# The nonparametric Fisher geometry and the chi-square process density prior

Andrew Holbrook[*,1], Shiwei Lan[2], Jeffrey Streets[3] and Babak Shahbaba[1]

[1]Department of Statistics, UC Irvine
[2]Department of Computing and Mathematical Sciences, Caltech
[3]Department of Mathematics, UC Irvine

November 14, 2017

## Abstract

It is well known that the Fisher information induces a Riemannian geometry on parametric families of probability density functions. Following recent work, we consider the nonparametric generalization of the Fisher geometry. The resulting nonparametric Fisher geometry is shown to be equivalent to a familiar, albeit infinite-dimensional, geometric object—the sphere. By shifting focus away from density functions and toward *square-root* density functions, one may calculate theoretical quantities of interest with ease. More importantly, the sphere of square-root densities is much more computationally tractable. This insight leads to a novel Bayesian nonparametric density estimation model. We construct the $\chi^2$-process density prior by modeling the square-root density with a restricted Gaussian process prior. Inference over square-root densities is fast, and the model retains the flexibility characteristic of Bayesian nonparametric models. Finally, we formalize the relationship between spherical HMC in the infinite-dimensional limit and standard Riemannian HMC.

## 1 Introduction

The Fisher information—and the geometry it induces—has been one of the unequivocal success stories of geometry in statistics. Building on recent work, we extend the Fisher geometry beyond parametric statistical models and show that the resulting geometry is equivalent to that of the infinite-dimensional sphere. The purpose of this paper is to bring attention to this new perspective and to demonstrate its theoretical and methodological consequences. As an application, we introduce the $\chi^2$-process density prior, a flexible nonparametric model for Bayesian density estimation that admits fast computation while requiring minimal assumptions.

The Fisher information matrix is canonical in statistics: it is rooted in information theory [1]; it appears in Jeffrey's prior of Bayesian analysis [2]; and it plays a central role in Bayesian and Frequentist asymptotics [3]. Fisher advocated the importance of the information matrix in maximum likelihood estimation [4]. Fisher's student, Rao, was the first to place the information matrix in a differential geometric context [5]. Since then, the differential geometric implications for parametric statistical models have been the subject of extensive inquiry [6]. Recently, a number of researchers have drawn connections between the Fisher geometry and the geometry of the infinite sphere [7]–[12]. Much of this work has been in the area of shape analysis and has focused on using the Fisher geometry to measure distance between probability densities. Bayesian uses for the nonparametric Fisher geometry were featured in [8], where Bayesian variational inference was accomplished by minimizing the Fisher distance, and in [10], where the nonparametric Fisher geometry was used for sensitivity analysis of Bayesian models. Here, we focus on fully Bayesian nonparametric inference, including the generation of posterior samples using Hamiltonian Monte Carlo (HMC). In contrast to recent research, the geodesics associated with the nonparametric Fisher geometry are used to efficiently explore the MCMC state space and *not* to measure or minimize the distance between density functions.

---

[*]aholbroo@uci.edu

This paper, and other recent research in the Fisher geometry, builds on the sub-field of square-root density estimation. [13] used a wavelet basis to estimate the square-root density by effectively fitting the curve and then normalizing a sparse collection of wavelet coefficients, and [14] introduced a Bayesian follow-up to this work. Recently, [15] used Riemannian geometry to fit a square-root density model, but did not make any connections to the Fisher geometry. More recently [12] performed square-root density estimation for object recognition using minimum description length as fitting criterion and used the nonparametric Fisher geometry to obtain a closed-form expression of this criterion.

In this paper, we focus on the application of the nonparametric Fisher geometry to Bayesian inference for probability densities. While the density function is the object of interest, we instead model the square-root density function, that is, the function the square of which integrates to unity. We take a Bayesian nonparametric approach and endow the square-root density with a Gaussian process (GP) prior [16], [17] multiplied by a Dirac measure limiting its support to the infinite-dimensional sphere. In order to maintain this restriction, it is useful to use the Karhunen-Loève (K-L) expansion [18] of the GP prior as opposed to its kernel representation. Every GP with bounded second moment may be represented in terms of the eigenfunction expansion of its covariance operator, but this (the K-L) expansion is only explicitly known for a few classes of GPs [18]. Still, the K-L expansion has seen much recent success in the realm of Bayesian inverse problems [19], [20] and has been featured in infinite-dimensional HMC and infinite manifold HMC ($\infty$-mHMC) [21]. The proposed application of the K-L expansion to model the square-root density is unprecedented and offers a probabilistic interpretation to the use of basis expansions for density estimation.

Due to the orthonormality of the eigenfunction basis, the restriction to the (uncountably) infinite-dimensional sphere translates to a restriction to the (countably) infinite-dimensional sphere for the eigenvalues of the GP. Then, following the precedent set in [21], the K-L expansion is truncated and the object of inference is reduced to the posterior distribution of a finite number of K-L coefficients restricted to a finite sphere. This computation is quick and easy using either spherical HMC [22]. Thanks to the basis representation, computational complexity scales linearly with the number of data points, as opposed the cubic rate of the GP density sampler [23]. Moreover, we show that—in the square-root density estimation context—spherical HMC corresponds to Riemannian HMC in the infinite-dimensional limit.

Squaring the GP square-root density prior gives a $\chi^2$-process [cf. 24] density prior. We illustrate the use of this prior for a number of problems. The model is flexible and its posterior draws provide plausible realizations of the uncertainty inherent in the density estimation problem. Besides a recent application to Bayesian quadrature [25], we are unaware of statistical applications for the $\chi^2$-process and are therefore pleased to present its novel application to Bayesian density estimation.

The contributions of this paper are as follows:

- we review a nonparametric generalization of the Fisher geometry and show its relationship to the infinite-dimensional ($L^2$) sphere, the space of square-root density functions;

- we derive the geodesics on the $L^2$ sphere and use these geodesics to formalize the relationship between Riemannian HMC and infinite-dimensional spherical HMC;

- focusing on Bayesian nonparametric density estimation, we demonstrate the practical benefits to modeling the *square-root* density function. The resulting $\chi^2$-process density prior performs well for a variety of problems and is efficiently computed using spherical HMC.

The rest of the paper is organized in the following way. In Section 2 we review the parametric Fisher geometry, present a nonparametric extension of the Fisher geometry, and derive key results by relating this geometry to the infinite-dimensional sphere. Section 3 presents the $\chi^2$-process density prior along with some necessary tools, such as the Karhunen-Loève expansion. In Section 4, we discuss efficient Bayesian inference for the model and relate Riemannian HMC to infinite-dimensional spherical HMC. Empirical results are presented in Section 5. Finally, in Section 6 we discuss model limitations and possible extensions.

## 2 The nonparametric Fisher geometry

### 2.1 The parametric Fisher geometry

Given data $x$ in domain $\mathcal{D}$, it is often useful to specify a probabilistic model $S = \{p_\theta = p(x, \theta) \,|\, \theta = [\theta^1, \ldots, \theta^p]\}$, where $\theta$ is a vector parameterizing the model and taking values in the continuous parameter space $\Theta$. Then at any point $\theta \in \Theta$, the Fisher information is the expectation of the negative log-likelihood Hessian:

$$\mathcal{I}(\theta) = -\mathrm{E}_x\Big(\frac{\partial^2 \ell(\theta)}{\partial \theta \partial \theta^T}\Big) = -\int_\mathcal{D} \frac{\partial^2 \ell(\theta)}{\partial \theta \partial \theta^T}\, p(x|\theta)\, \mu(dx)\,, \tag{1}$$

where $\ell(\theta) = \log p(x|\theta)$. In the language of optimization, the Fisher information encodes second-order functional information about $\ell(\theta)$. This fact explains the use of the Fisher information as a gradient preconditioning matrix in both (the Frequentist) Fisher scoring [26] and (the Bayesian) Riemannian HMC [27]. The Fisher information may also be written as the expected outer product of the score vector $\partial \log p(x|\theta)/\partial \theta$:

$$\mathcal{I}(\theta) = \mathrm{E}_x\Big(\big(\frac{\partial \ell(\theta)}{\partial \theta}\big)\big(\frac{\partial \ell(\theta)}{\partial \theta}\big)^T\Big) = \int_\mathcal{D} \big(\frac{\partial \ell(\theta)}{\partial \theta}\big)\big(\frac{\partial \ell(\theta)}{\partial \theta}\big)^T p(x|\theta)\, \mu(dx)\,. \tag{2}$$

The Fisher information is symmetric positive definite at any point $\theta \in \Theta$. Taking note of this fact, Rao [5] interpreted the Fisher information matrix as a Riemannian metric tensor, i.e. a smoothly varying, symmetric positive definite matrix defined over the parameter space $\Theta$. In this way, the Fisher information matrix induces a Riemannian metric $g_\theta(\cdot, \cdot)$ over $\Theta$ satisfying

$$g_\theta(\ell_i, \ell_j) = \mathcal{I}_{ij}(\theta)\,, \quad \text{and} \quad g_\theta(\psi, \phi) = \sum_{i,j} \psi^i \phi^j \mathcal{I}_{ij}(\theta) \tag{3}$$

for $\ell_i = \partial \ell(\theta)/\partial \theta_i$, $\psi = \sum_{k=1}^p \psi^k \ell_k$ and $\phi = \sum_{k=1}^p \phi^k \ell_k$. Hence, the Fisher information may be thought of as inducing a non-trival geometry on the otherwise Euclidean parameter space $\Theta$. There has been much inquiry into the nature of the parametric Fisher geometry. Efron used the Fisher geometry to prove the second-order efficiency of the MLE for exponential family models [28], and Amari and Nagaoka [6] has constructed a body of work around the Fisher geometry and its dual connections. More recently, Girolami and Calderhead [27] successfully used the Fisher geometry to guide the Hamiltonian flow of their Riemannian HMC. In this paper, we take another tact by generalizing the notion of the Fisher geometry to nonparametric models.

### 2.2 Beyond parametric models

We consider probability distributions over smooth manifolds $\mathcal{D}$, of which $\mathcal{D} \cong \mathbb{R}^d$ is a special case. Having fixed a background measure $\mu$, let

$$\mathcal{P} := \Big\{p : \mathcal{D} \to \mathbb{R} \,\Big|\, p \geq 0, \int_\mathcal{D} p(x)\, \mu(dx) = 1\Big\} \tag{4}$$

be the space of probability density functions over $\mathcal{D}$. That is, $\mathcal{P}$ is the set of Radon-Nikodym derivatives of probability measures that are absolutely continuous with respect to $\mu$. The following construction is agnostic to whether $\mu$ is the Lebesgue measure over $\mathcal{D} = \mathbb{R}^d$ or the Hausdorff measure over a general Riemannian manifold $\mathcal{D} = \mathcal{M}$.

We deal with the space $\mathcal{P}$ and do *not* fix a parametric model. Instead we give $\mathcal{P}$ the structure of an infinite dimensional (formal) Riemannian manifold. First, we think of it as a smooth manifold. Observe that for a given $p \in \mathcal{P}$, the tangent space can be identified with

$$T_p \mathcal{P} := \Big\{\phi \in C^\infty(\mathcal{D}) \,\Big|\, \int_\mathcal{D} \phi(x)\, \mu(dx) = 0\Big\}\,. \tag{5}$$

This identification arises when one differentiates the unit measure condition on probability density functions. That is, for a smooth curve $p_t : (-\epsilon, \epsilon) \to \mathcal{P}$ satisfying $dp_t/dt|_{t=0} = \phi$, we have

$$0 = \frac{d}{dt} \int_{\mathcal{D}} p_t(x)\, \mu(dx)\Big|_{t=0} = \int_{\mathcal{D}} \frac{dp_t}{dt}(x)\, \mu(dx) = \int_{\mathcal{D}} \phi(x)\, \mu(dx). \tag{6}$$

Now that we have a smooth manifold and an associated tangent space, we may define a Riemannian metric, i.e. a smoothly varying, symmetric, non-degenerate, bilinear function $g(\cdot,\cdot)_p : T_p\mathcal{P} \times T_p\mathcal{P} \to \{0\} \cup \mathbb{R}^+$. Riemannian metrics are useful for developing a notion of distance on a manifold that does not depend on any embedding in Euclidean space. One may define uncountably many metrics on a general manifold, but we are interested in a generalization of the parametric Fisher information metric.

**Definition 1.** *([7], [11]) Given $\mathcal{D}$, the* nonparametric Fisher information metric *on $\mathcal{P}(\mathcal{D})$[1] is*

$$g_F(\phi, \psi)_p := \int_{\mathcal{D}} \frac{\phi(x)\psi(x)}{p(x)} \mu(dx). \tag{7}$$

This metric is a consistent generalization of the parametric Fisher information metric. To see this, consider the parametric model $p(x|\theta)$, with $\theta$ as a vector. Then each element $\theta_i$ of $\theta$ defines a curve $\Theta_i \to \mathcal{P}$, where $\Theta_i$ is a slice of $\Theta$, and

$$\mathcal{I}_{ij}(\theta) = \int_{\mathcal{D}} \ell_i \ell_j\, p(x|\theta)\mu(dx) = \int_{\mathcal{D}} \frac{p_i(x|\theta)}{p(x|\theta)} \frac{p_j(x|\theta)}{p(x|\theta)}\, p(x|\theta)\mu(dx) = \int_{\mathcal{D}} \frac{p_i(x|\theta)p_j(x|\theta)}{p(x|\theta)} \mu(dx). \tag{8}$$

Here, we have adopted the shorthand $p_i(x|\theta) = \partial p(x|\theta)/\partial \theta_i$. Expressed in a more invariant fashion, interpreting a model as a map $\theta : \Theta \to \mathcal{P}$, one has that the parametric Fisher metric is induced by the nonparameteric Fisher metric, i.e.

$$\theta^* g_F = g_\theta. \tag{9}$$

In what follows we make a nontrivial change of variables suggested by this geometric picture which provides various theoretical and computational simplifications. In particular, for various reasons the manifold $\mathcal{P}$ equipped with Riemannian metric (7) is not particularly easy to deal with. In order to calculate geometric quantities of interest (e.g. geodesics, distances), we shift focus to the $L^2$ unit sphere, i.e. the space of square-root density functions

$$\mathcal{Q} := \left\{ q : \mathcal{D} \to \mathbb{R} \,\Big|\, \int_{\mathcal{D}} q(x)^2 \mu(dx) = 1 \right\}. \tag{10}$$

This space, which is identified with $\mathcal{P}$ by a simple transformation indicated below, provides a much simpler backdrop for calculations. This infinite-dimensional $L^2$ sphere is a surprisingly familiar object. Its tangent spaces and geodesics are formally the exact same as those of the finite dimensional sphere $\mathcal{S}^{n-1}$, the only difference being the replacement of the Euclidean inner product with the integral inner product of $L^2$:

$$\langle f, h \rangle_{L^2} = \int_{\mathcal{D}} f(x)h(x)\, \mu(dx). \tag{11}$$

Remarkably, this simpler space is isometric to the space of density functions equipped with the nonparameteric Fisher metric defined above.

**Lemma 1.** *The map $S : (\mathcal{P}, g_F) \to (\mathcal{Q}, \langle \cdot, \cdot \rangle_{L^2})$ defined by $S(p) := 2\sqrt{p}$ is a Riemannian isometry.*

---

[1] From the definition, the nonparametric Fisher metric can take on infinite values. It is possible to avoid this by limiting the space of interest to strictly-positive density functions or by bounding the metric at an arbitrarily large value. It is also possible to modify the definition of the tangent space to enforce tangent functions to equal 0 when their respective densities do.

*Proof.* We must show that $\langle S_*\psi, S_*\phi\rangle_{L^2} = g_F(\psi, \phi)_p$, where $S_*$ is the pushforward (or Jacobian) of $S$:

$$S_* = \frac{dS}{dp}(p) = \frac{d(2\sqrt{p})}{dp} = \frac{1}{\sqrt{p}}. \tag{12}$$

By direct computation,

$$\langle S_*\psi, S_*\phi\rangle_{L^2} = \int_{\mathcal{D}} (S_*\psi)(x)(S_*\phi)(x)\,\mu(dx) = \int_{\mathcal{D}} \frac{\psi(x)}{\sqrt{p(x)}} \frac{\phi(x)}{\sqrt{p(x)}} \mu(dx) \tag{13}$$

$$= \int_{\mathcal{D}} \frac{\psi(x)\phi(x)}{p(x)} \mu(dx) = g_F(\psi, \phi)_p.$$

$\square$

In the remainder of this section we present a few basic results regarding the nonparametric Fisher geometry, working with the $L^2$ sphere model and transferring results to the traditional Fisher geometry. We note that investigations of the nonparameteric Fisher information have independently appeared in [7]–[12]. We reproduce some fundamental aspects of this geometry relevant to Theorem 1 (Section 4.1) for convenience. To begin we observe how to describe the tangent space to $\mathcal{Q}$.

**Lemma 2.** *Given $q \in \mathcal{Q}$, one has that*

$$T_q\mathcal{Q} := \left\{ f : \mathcal{D} \to \mathbb{R} \mid \int_{\mathcal{D}} q(x)f(x)\,\mu(dx) = 0 \right\}. \tag{14}$$

*Proof.* If $q_t : (-\epsilon, \epsilon) \to \mathcal{Q}$ denotes a path in $Q$ satisfying $dq_t/dt|_{t=0} = f$, then the unit integration constraint on $p = q^2$ means

$$0 = \frac{d}{dt}\int_{\mathcal{D}} q_t(x)^2 \mu(dx)\Big|_{t=0} = 2\int_{\mathcal{D}} q_0(x)\frac{dq}{dt}(x)\Big|_{t=0}\mu(dx) = 2\int_{\mathcal{D}} q_0(x)f(x)\mu(dx). \tag{15}$$

$\square$

We next solve one version of the geodesic problem on $\mathcal{P}$. In particular we consider an initial point and velocity and solve for continuing the geodesic in that direction. We will exploit the isometry between $\mathcal{P}$ and $\mathcal{Q}$ and solve first in $\mathcal{Q}$.

**Lemma 3.** *Given $q_0 \in \mathcal{Q}$ and $f \in T_q\mathcal{Q}$ a unit vector, the geodesic with initial condition $q_0$ and velocity $f$ exists on $(-\infty, \infty)$ and takes the form*

$$q_t = q_0 \cos t + f \sin t. \tag{16}$$

*Proof.* First we derive the geodesic equation in $\mathcal{Q}$. One conceptual method for obtaining this, exploiting the spherical structure of $\mathcal{Q}$, is to first observe that if $q_t$ is a path in $\mathcal{Q}$ and $a_t \in T_{q(t)}\mathcal{Q}$ is a tangent vector along the curve, the Fisher geometry induces a covariant derivative along the path via

$$\frac{D}{\partial t}a = \dot{a} - q\int_{\mathcal{D}} \dot{a}q, \tag{17}$$

which is manifestly the time derivative of the family $a_t$ projected to the tangent space at $q_t$, as expected. For a curve $q_t$ to be a geodesic, it should have zero acceleration, i.e.

$$0 = \frac{D}{\partial t}\dot{q} = \ddot{q} - q\int_{\mathcal{D}} \ddot{q}q. \tag{18}$$

However, using that $\int_\mathcal{D} q_t(x)^2 \mu(dx) = 1$ for all $q$ and differentiating twice in $t$, one sees that this is equivalent to

$$\ddot{q} + q \int_\mathcal{D} \dot{q}^2 = 0, \tag{19}$$

which we now take as the geodesic equation in $\mathcal{Q}$. Another method for deriving this equation is to solve for which curves are critical points for the length functional with fixed endpoints.

Now, to solve this equation in our setting, first let us observe that since $f \in T_{q_0}\mathcal{Q}$, by Lemma 2 we have

$$\int_\mathcal{D} q_0 f = 0. \tag{20}$$

Using this and the fact that $f$ is a unit vector we compute

$$\begin{aligned}
\frac{d}{dt} \int_\mathcal{D} \dot{q}^2 &= 2 \int_\mathcal{D} \ddot{q}\dot{q} \\
&= 2 \int_\mathcal{D} [-q_0 \cos t - f \sin t][-q_0 \sin t + f \cos t] \\
&= 2 \int_\mathcal{D} [q_0^2 - f^2] \cos t \sin t \\
&= 0.
\end{aligned} \tag{21}$$

Thus $\int_M \dot{q}^2 = \int_\mathcal{D} f^2 = 1$. We then simply observe the ODE

$$\ddot{q} = -q, \tag{22}$$

and it is clear that $q$ satisfies (19), and so the lemma follows. □

We now translates this result into a corresponding one for geodesics in $\mathcal{P}$.

**Lemma 4.** *Given $p_0 \in \mathcal{P}$ and $f \in T_p\mathcal{P}$ a unit vector, the geodesic with initial condition $p_0$ and initial velocity $f$ exists on $(-\infty, \infty)$, and takes the form*

$$p_t = \left(\sqrt{p_0} \cos t + \frac{f}{2\sqrt{p_0}} \sin t\right)^2. \tag{23}$$

*Proof.* We use Lemma 3 and reinterpret the geodesic equation in terms of square-roots. In this formalism the initial condition is $q_0 = \sqrt{p_0}$ and the initial velocity is

$$\frac{d}{dt}q = \frac{d}{dt}\sqrt{p} = \frac{f}{2\sqrt{p_0}} = \frac{f}{2q_0}.$$

□

These basic lemmas show the advantage of working in $\mathcal{Q}$, yielding a conceptual derivation of the geodesic equation. These lemmas will be used to prove Theorem 1 in Section 4.1. As we will see below, not only is the $L^2$ sphere $\mathcal{Q}$ more theoretically tractable, it also turns out to be more computationally tractable. In the following sections, we take advantage of these two kinds of tractability to construct a Bayesian nonparametric model on $\mathcal{Q}$ and use it for an application in density estimation.

# 3 The chi-square process density prior

In this section, we transition from the theoretical to the applied aspects of the nonparametric Fisher geometry. We find that the square-root representation $q = \sqrt{p}$ is of use practically as well as theoretically. Here we focus on its natural application for density estimation.

A good density estimate places more mass where there is more data but takes the finite nature—and the uncertainty that comes with it—of that data into account. Bayesian non-parametric density estimation effects this balance: non-parametric models give flexibility, while the Bayesian prior contributes regularization. These methods model the data generating distribution as a random function, itself drawn from a specified stochastic process. Dirichlet processes mixture models (DPMMs) convolve the Dirichlet process with a smooth distribution, in effect constructing an infinite mixture model [29]. More recently, [23] proposed a new method, called Gaussian Process Density Sampler (GPDS), offering a similar amount of flexibility as the DPMM but having an arguably simpler framework. Nonetheless, inference for DPMMs requires an advanced Gibbs sampling routine [30], and inference for the GPDS requires exchange sampling to handle the unit-integral restriction on the GP model [23]. In contrast, the model we propose here can be computed using generic spherical HMC [22] or geodesic Monte Carlo [31] algorithms. Further, we take a different approach from other Bayesian nonparametric density models by modeling the square-root density function instead. In the previous section, theoretical results for the nonparametric Fisher geometry were easier to obtain by first obtaining the corresponding results on the $L^2$ sphere and then translating the results to the Fisher geometry. This theme continues in application, where we show that Bayesian density estimation can be much easier when one shifts focus to the sphere of square-root densities. We place a GP prior on the square-root of the probability density function. This amounts to a $\chi^2$-process prior on the density function itself.

Suppose we want to attribute a smooth density function to observed data $x_1, \ldots, x_n$ on finite domain $\mathcal{D} \subset \mathbb{R}^d$ and recall the definitions (from Section 2) of the space of density functions and the space of square-root density functions:

$$\mathcal{P} := \left\{ p : \mathcal{D} \to \mathbb{R} \mid p \geq 0, \int_\mathcal{D} p(x)\,\mu(dx) = 1 \right\} \quad \text{and} \quad \mathcal{Q} := \left\{ q : \mathcal{D} \to \mathbb{R} \mid \int_\mathcal{D} q(x)^2\,\mu(dx) = 1 \right\}, \quad (24)$$

respectively. We want to find a suitable element $p(\cdot) \in \mathcal{P}(\mathcal{D})$, the space of functions over domain $\mathcal{D}$. Although this space contains the functions of interest, we opt to deal with the space $\mathcal{Q}$ of square-root densities instead. As stated in the prior section, $\mathcal{Q}$ is the unit sphere in the infinite-dimensional Hilbert space $L^2(\mathcal{D})$. We model the square-root density with a GP prior (or a Gaussian measure in $L^2$) multiplied by the Dirac measure restricting the function to the unit sphere:

$$q \sim \mathcal{GP} \times \delta_q(\mathcal{Q}). \quad (25)$$

It turns out that it is much easier to enforce the constraint given by Dirac measure $\delta_q(\mathcal{Q})$ than it is to enforce the corresponding constraint $\delta_p(\mathcal{P})$ (as is done for the GPDS). To do so, however, we do not represent the GP prior using its kernel representation as is commonly done in the literature [32]. We opt instead to represent $q$ in terms of the eigenvalues and orthonormal eigenfunctions of its covariance operator.

## 3.1 The Karhunen-Loève representation

In order to tractably enforce the constraint $\delta_q(\mathcal{Q})$ in (25), it is helpful to write $q$ as a function (or linear sum of functions) for which we know the values of both

$$\int_\mathcal{D} q(x)\mu(dx) \quad \text{and} \quad \int_\mathcal{D} q(x)^2\mu(dx). \quad (26)$$

This condition is satisfied by representing random function $q$ as a linear combination of orthonormal basis functions. The K-L representation [18] provides a canonical way of doing so and thus links our fully probabilistic approach to other square-root density methods that rely on a basis [13]–[15]. Let $u(\cdot) \sim \mathcal{GP}(0, K(\cdot))$ be a mean zero Gaussian process over domain $\mathcal{D}$ with covariance operator $K(\cdot)$. Then $u$ admits a K-L expansion of the form

$$u(\cdot) = \sum_{i=1}^{\infty} u_i \, \phi_i(\cdot), \qquad u_i \stackrel{ind}{\sim} N(0, \lambda_i^2), \tag{27}$$

where the $\lambda_i$s and the $\phi_i$s are respectively the eigenvalues and eigenfunctions of operator $K$. That is to say, they satisfy

$$K(\phi_i)(x') = \int k(x, x')\phi_i(x)\mu(dx) = \lambda_i \phi_i(x') \tag{28}$$

where $k(\cdot, \cdot)$ is the usual covariance kernel. The eigenvalues are decreasing and their sum-of-squares is finite: $\lambda_{i+1} < \lambda_i$, $\sum_{i=1}^{\infty} \lambda_i^2 < \infty$. Finally, the eigenfunctions form an orthonormal basis of $L^2$:

$$\int \phi_i(x)\phi_j(x)\mu(dx) = 0, \quad \text{and} \quad \int \phi_i^2(x)\mu(dx) = 1. \tag{29}$$

In this paper, we model $q$ as belonging to the Matérn class of GPs. For the Matérn class, a closed-form orthonormal basis may be obtained from the eigenfunctions of the Laplacian [21], [33]. The covariance operator is given by

$$K = \sigma^2 (\alpha - \Delta)^{-s}, \tag{30}$$

where $\alpha$ and $\sigma^2$ are positively constrained scale parameters, $s$ is a smoothness parameter, and $\Delta$ is the Laplacian $\sum_{i=1}^{d} \partial_i^2$. The eigenvalues and eigenfunctions corresponding to this covariance operator depend on the area and dimensionality of domain $\mathcal{D}$ and are presented in Section 5 below. It should be noted that the decision to use the Matérn class is entirely dictated by ease of computation and does not preclude other classes of GP from being used in future applications.

## 3.2 The model

The proposed density model is Bayesian nonparametric, i.e. we place a prior distribution on a set of functions and eschew a restrictive parametric form. Given data $x = (x_1, \cdots, x_N) \in \mathcal{D}$, we obtain a posterior distribution, which is itself a distribution over the same set of functions and is absolutely continuous with respect to the specified prior distribution. As stated above, the prior $\pi(q)$ on square-root density $q \in \mathcal{Q}$ is a GP multiplied by the Dirac measure on the $L^2$ sphere. Following (27), the prior for $q$ and the likelihood of the data $x$ given $q$ are given by

$$\pi(q) \propto \delta_q(\mathcal{Q}) \prod_{i=1}^{\infty} \exp\left(-q_i^2/(2\lambda_i^2)\right), \quad \text{and} \quad \pi(x|q) = \prod_{n=1}^{N} q^2(x_n), \tag{31}$$

since $q$ is the square-root density. This prior can also be interpreted as arising from an infinite-dimensional Bingham distribution on the coefficients [34]. The posterior distribution on $q$ is then given by

$$\pi(q|x) = \frac{\pi(x|q)\,\pi(q)}{\int_{\mathcal{Q}} \pi(x|q)\,\pi(q)\,dq} \propto \pi(q) \prod_{n=1}^{N} q^2(x_n). \tag{32}$$

Suppressing the Dirac measure, the log-posterior given data $x_{1:N}$ may be written in terms of the K-L expansion (27) of $q$:

$$\log \pi(q|x) \propto \sum_{n=1}^{N} \log q(x_n)^2 - \frac{1}{2}\sum_{i=1}^{\infty} q_i^2/\lambda_i^2 \tag{33}$$

$$= 2\sum_{n=1}^{N} \log |q(x_n)| - \frac{1}{2}\sum_{i=1}^{\infty} q_i^2/\lambda_i^2$$

$$= 2\sum_{n=1}^{N} \log |\sum_{i=1}^{\infty} q_i \phi_i(x_n)| - \frac{1}{2}\sum_{i=1}^{\infty} q_i^2/\lambda_i^2 \,.$$

By modelling the square-root density $q$ with a GP prior, we model the density function $p$ with a $\chi^2$-process prior. Modeling the density $p$ as a $\chi^2$-process, we automatically enforce the non-negativity requirement for probability density functions. On the other hand, $\chi^2$-processes are not restricted to have unit integrals. We therefore rely on a geometric HMC inference scheme to restrict proposals to the $L^2$ sphere. This is discussed in the following section.

## 4 Inference

Inference for the $\chi^2$-process density model is relatively straightforward and amenable to advanced HMC methods. In Section 4.1, we show that, in this context, infinite-dimensional spherical HMC is equivalent to Riemannian HMC using the parametric Fisher information. In practice, we follow Beskos, Girolami, Lan, et al. [21] and truncate[2] the K-L expansion of the GP square-root density prior for an integer $I$ using truncation operator $T_I$:

$$T_I\big(q(x)\big) = T_I\Big(\sum_{i=0}^{\infty} q_i \,\phi_i(x)\Big) = \sum_{i=0}^{I} q_i \,\phi_i(x) \,. \tag{34}$$

Due to the orthonormality of the basis $\phi_i$, the unit integral constraint on $T_I(q)^2$ translates directly to a spherical constraint on the random coefficients $q^I = (q_0, \cdots, q_I)$. That is,

$$1 = \int_{\mathcal{D}} T_I\big(q(x)\big)^2 \mu(dx) = \int_{\mathcal{D}} \Big(\sum_{i=0}^{I} q_i \,\phi_i(x)\Big)^2 \mu(dx) = \sum_{i=0}^{I} q_i^2 \int \phi_i(x)^2 \mu(dx) = \sum_{i=0}^{I} q_i^2 \tag{35}$$

where the penultimate equality is given by the orthogonality of the basis elements and the last equality is on account of the basis elements being normal. Thus, inference can be performed over the coefficients $q^I$ by using spherical HMC [22] on the sphere $\mathcal{S}^I$. Both of these methods augment the state space with an auxiliary velocity variable $v$ (satisfying $v^T q^I = 0$) and simulate from a Hamiltonian system by splitting [36] the Hamiltonian of interest ($H$) into two Hamiltonians ($H^1 + H^2$):

$$H(q^I, v) = -\log \pi(q^I) + \frac{1}{2}G(q^I) + \frac{1}{2}v^T v \tag{36}$$

$$H^1(q^I, v) = -\log \pi(q^I) + \frac{1}{2}G(q^I)$$

$$H^2(q^I, v) = \frac{1}{2}v^T v \,.$$

---

[2] We note that one may conceivably place a prior on the truncation index $I$ and thus avoid having to choose the number of eigenfunctions. This would provide for an interesting extension of the model presented here, but would necessitate new MCMC techniques that enable the change of model dimensionality (e.g. reversible jump MCMC [35]) *while maintaining manifold constraints*. Hence, we leave this for future work.

Here $\pi$ is the posterior distribution and $G$ is the canonical Riemann tensor for the sphere [22]. Simulating from $H^1$ involves a small perturbation of the velocity by the gradient of $H^1$ with respect to $q^I$; simulating $H^2$ involves moving along the sphere's geodesics in the direction $v$. This last fact is relevant to the discussion of the following section.

Spherical HMC requires the gradient of the log-posterior with respect to the coefficients. Elementwise, this is given by

$$\frac{\partial}{\partial q_j} \log \pi(q^I|x) = 2 \sum_{n=1}^{N} \frac{\partial}{\partial q_i} \log |\sum_{i=1}^{I} q_i \phi_i(x_n)| - \frac{1}{2} \frac{\partial}{\partial q_j} \sum_{i=1}^{I} q_i^2 / \lambda_i^2 \qquad (37)$$

$$= 2 \sum_{n=1}^{N} \frac{\phi_j(x_n)}{\sum_{i=1}^{I} q_i \phi_i(x_n)} - q_j / \lambda_j^2 \,.$$

The Markov chain may be initialized using Newton's method on the sphere (see Appendix A).

Since the values of the eigenfunctions at the observations may be precomputed, the main computational burden is in the summations involved in the evaluation of the log-posterior and its gradient. Since in practice $I \ll N$, these computations are $O(N)$, where $N$ is the number of data points. This is orders faster than the $O(N^3)$ computations required to perform inference for the GPDS [23].

### 4.1 Inference in the limit

We note that both spherical HMC uses geodesic flows on the finite dimensional sphere to propose new Markov chain states. Since these flows are formally equivalent to the geodesic flows on the $L^2$ sphere (see Section 2) and since the natural geometry on $L^2$ is equivalent to the nonparametric Fisher geometry, it is worth asking whether these inference schemes are adapted to the nonparametric Fisher geometry in a similar way to Riemannian HMC's adaptation to the parametric Fisher geometry[3].

Indeed this is the case, and it is a simple consequence of Lemma 1 and the isometric relationship between square-integrable functions and square-summable sequences induced by any orthonormal basis $\{\phi_i\}_{i=1}^{\infty}$ with completion $L^2$. Denote the space of square-summable sequences and its sphere

$$\ell^2 = \left\{ q = \{q_i\}_{i=1}^{\infty} \,\middle|\, \langle q, q \rangle_{\ell^2} = \sum_{i=1}^{\infty} q_i^2 < \infty \right\} , \quad \mathcal{S}^{\infty} = \left\{ q \in \ell^2 \,\middle|\, \langle q, q \rangle_{\ell^2} = \sum_{i=1}^{\infty} q_i^2 = 1 \right\} . \qquad (38)$$

Then it follows from the orthonormality of $\{\phi_i\}_{i=1}^{\infty}$ that $(L^2, \langle \cdot, \cdot \rangle_{L^2}) \cong (\ell^2, \langle \cdot, \cdot \rangle_{\ell^2})$, since for any arbitrary function $q = q(\cdot) \in L^2$,

$$\langle q, q \rangle_{L^2} = \int q(x)^2 \mu(dx) = \int \Big( \sum_{i=1}^{\infty} q_i \phi_i(x) \Big)^2 \mu(dx) = \sum_{i=1}^{\infty} q_i^2 = \langle q, q \rangle_{\ell^2} \,. \qquad (39)$$

It is an immediate result that the respective spheres are also isometric, i.e. $(\mathcal{Q}, \langle \cdot, \cdot \rangle_{L^2}) \cong (\mathcal{S}^{\infty}, \langle \cdot, \cdot \rangle_{\ell^2})$, and hence, by Lemma 1, the following result holds.

**Lemma 5.** *Given an orthonormal basis for $L^2$, the space of density functions equipped with the Fisher metric is isometric to the sphere $\mathcal{S}^{\infty}$ with its natural Euclidean metric, i.e. $(\mathcal{P}, g_F(\cdot, \cdot)) \cong (\mathcal{S}^{\infty}, \langle \cdot, \cdot \rangle_{\ell^2})$.*

Our goal is to show that spherical HMC is adapted to the nonparametric Fisher geometry in the infinite-dimensional limit. Given that the geodesic paths followed by spherical HMC converge to geodesics on $\mathcal{S}^{\infty}$, Lemma 5 will imply that these paths correspond to geodesics on $(\mathcal{P}, g_F(\cdot, \cdot))$.

**Lemma 6.** *Geodesic flows on the finite sphere $\mathcal{S}^{I-1}$ converge to geodesic flows on the infinite-dimensional sphere $\mathcal{S}^{\infty}$ as $I \to \infty$.*

---
[3]By Riemannian HMC, we mean Riemannian HMC where the Riemannian metric is the finite Fisher metric, as this is the most common usage. We note that it is theoretically possible to use other metrics [31], [37].

*Proof.* For any point $q \in \mathcal{S}^\infty$, let $q^I \in \mathcal{S}^{I-1}$ be vector obtained by applying the truncation operator to $q$ and then normalizing:

$$q^I = \frac{T_I(q)}{\|T_I(q)\|} = \frac{(q_1, \ldots, q_I)^T}{\sqrt{(q_1, \ldots, q_I)(q_1, \ldots, q_I)^T}}. \tag{40}$$

Similarly, for any vector in the tangent space to $S^\infty$

$$v \in T_q \mathcal{S}^\infty = \left\{ v \in \ell^2 \,\big|\, \langle v, q \rangle_{\ell^2} = \sum_{i=1}^\infty q_i v_i = 0 \right\} \tag{41}$$

let $v^I \in T_{q^I} \mathcal{S}^{I-1}$ be the $I$-dimensional vector obtained by truncating $v$, projecting onto the tangent space $T_{q^I} \mathcal{S}^{I-1}$, and scaling such that $\|v\|_{\ell^2} = \|v^I\|$ (where $\|\cdot\|$ is the Euclidean norm):

$$\tilde{v}^I = T_I(v) - q^I \langle q^I, T_I(v) \rangle_{\ell^2}, \quad \text{and} \quad v^I = \tilde{v}^I \frac{\|v\|_{\ell^2}}{\|\tilde{v}^I\|}. \tag{42}$$

It follows from the definition of truncation (Equation (34)) that $q^I \to q$ and $v^I \to v$ with respect to $\langle \cdot, \cdot \rangle_{\ell^2}$ as $I \to \infty$.

Next, let $t \mapsto (q(t), v(t))$ be the geodesic flow on $\mathcal{S}^\infty$ with initial position $q_0 = q(0)$ and initial velocity $v_0 = v(0) \in T_{q_0} \mathcal{S}^\infty$. Let $t \mapsto (q^I(t), v^I(t))$ be the analogous flow on the tangent bundle $T\mathcal{S}^{I-1}$, where $q_0^I$ and $v_0^I$ are obtained from $q_0$ and $v_0$ following Formulas (41) and (42), respectively. Denote the distance between flows at time $t$

$$f(t) = \|q_t - q_t^I\|_{\ell^2}^2 + \|\dot{q}_t - \dot{q}_t^I\|_{\ell^2}^2. \tag{43}$$

Our goal is to show that

$$\lim_{I \to \infty} \int_0^T f(t)\, dt = 0, \tag{44}$$

for any finite $T$, and hence that geodesic flows on the finite sphere converge to those on $\mathcal{S}^\infty$. Begin by bounding $\dot{f}(t)$ by a constant times $f(t)$:

$$\frac{d}{dt} f(t) = 2 \left( \langle q_t - q_t^I, \dot{q}_t - \dot{q}_t^I \rangle_{\ell^2} + \langle \dot{q}_t - \dot{q}_t^I, \ddot{q}_t - \ddot{q}_t^I \rangle_{\ell^2} \right) \tag{45}$$
$$= 2 \left( \langle q_t - q_t^I, \dot{q}_t - \dot{q}_t^I \rangle_{\ell^2} + \langle \dot{q}_t - \dot{q}_t^I, -q_t \|\dot{q}_t\|_{\ell^2}^2 + q_t^I \|\dot{q}_t^I\|^2 \rangle_{\ell^2} \right).$$

Here, the second line follows from the geodesic formula. Noting that $\|\dot{q}_t\|_{\ell^2}^2 = \|\dot{q}_0\|_{\ell^2}^2$, $\|\dot{q}_t^I\|^2 = \|\dot{q}_0^I\|^2$, and that (by Equation (42)) $\|\dot{q}_0^I\|^2 = \|\dot{q}_0\|_{\ell^2}^2$, we get

$$\frac{d}{dt} f(t) = 2 \left( \langle q_t - q_t^I, \dot{q}_t - \dot{q}_t^I \rangle_{\ell^2} - \langle \dot{q}_t - \dot{q}_t^I, q_t - q_t^I \rangle_{\ell^2} \|\dot{q}_0\|_{\ell^2}^2 \right) \tag{46}$$
$$= 2 \left( 1 - \|\dot{q}_0\|_{\ell^2}^2 \right) \langle q_t - q_t^I, \dot{q}_t - \dot{q}_t^I \rangle_{\ell^2}.$$

We obtain our bounds by noting that

$$0 \leq \|q_t - q_t^I\|_{\ell^2}^2 - 2\langle q_t - q_t^I, \dot{q}_t - \dot{q}_t^I \rangle_{\ell^2} + \|\dot{q}_t - \dot{q}_t^I\|_{\ell^2}^2 \tag{47}$$
$$= f(t) - 2\langle q_t - q_t^I, \dot{q}_t - \dot{q}_t^I \rangle_{\ell^2} \tag{48}$$
$$= f(t) - \frac{\dot{f}(t)}{1 - \|\dot{q}_0\|_{\ell^2}^2},$$

and hence that

$$\frac{d}{dt} f(t) \leq \left(1 - \|\dot{q}_0\|_{\ell^2}^2\right) f(t) \,. \tag{49}$$

Integrating gives

$$f(t) \leq f(0) \, e^{t \, (1 - \|\dot{q}_0\|_{\ell^2}^2)} \,. \tag{50}$$

Since, by definition, $f(0) \to 0$ as $I \to \infty$, we have

$$\int_0^T f(t) \, dt \leq f(0) \int_0^T e^{t \, (1 - \|\dot{q}_0\|_{\ell^2}^2)} \, dt \tag{51}$$
$$= c \, f(0) \longrightarrow 0 \,.$$

Thus we have proven the convergence of geodesic flows on the finite sphere to those on $\mathcal{S}^\infty$. $\square$

We are now ready to connect Riemannian HMC and spherical HMC in the infinite-dimensional limit (where the latter is applied to the square-root density estimation problem). To make this relationship as clear as possible, we introduce a different (but equivalent) definition of a geodesic based on the calculus of variations (in contrast to the null acceleration definition from Lemma 3). Assume that two points $A$ and $B$ are close together in a small open set of Riemannian manifold $(\mathcal{M}, g(\cdot, \cdot))$. Let $\Gamma : [a, b] \times (-\epsilon, \epsilon) \to \mathcal{M}$ be a family of curves $\gamma_s : [a, b] \to \mathcal{M}$ satisfying $\gamma_s(a) = A$ and $\gamma_s(b) = B$ for all $s \in (-\epsilon, \epsilon)$. Then $\gamma$ is a geodesic if it minimizes the energy functional

$$E(\gamma) = \frac{1}{2} \int_a^b g_{\gamma(t)}\bigl(\dot{\gamma}(t), \dot{\gamma}(t)\bigr) \, dt \,, \quad \text{and thus satisfies} \quad \frac{d}{ds} E(\gamma_s) = 0 \,. \tag{52}$$

For a parametric family of distributions $\mathcal{P}_\theta$ equipped with the Fisher metric, the *parametric Fisher energy* takes the form

$$E(\theta) = \frac{1}{2} \int_a^b g_{\theta(t)}\bigl(\dot{\theta}(t), \dot{\theta}(t)\bigr)_F \, dt = \frac{1}{2} \int_a^b \nabla_\theta \ell(\theta(t))^T \, \mathcal{I}(\theta(t))^{-1} \, \nabla_\theta \ell(\theta(t)) \, dt \,, \tag{53}$$

where $\mathcal{I}(\theta)$ is the Fisher information, and $\ell(\theta) = \log p(\theta)$. On the other hand by Lemmas 1 and 5, the *nonparametric Fisher energy* for a family of curves in $\mathcal{P}$ takes the form

$$E(p) = \frac{1}{2} \int_a^b g_{p(t)}\bigl(\dot{p}(t), \dot{p}(t)\bigr)_F \, dt = \frac{1}{2} \int_a^b \langle \dot{q}(t), \dot{q}(t) \rangle_{L^2} \, dt = \frac{1}{2} \int_a^b \langle \dot{q}(t), \dot{q}(t) \rangle_{\ell^2} \, dt \tag{54}$$

where $q = \sqrt{p} = \sum_{i=1}^\infty q_i \phi_i(\cdot)$.

**Theorem 1.** *Let $q(\cdot) = \sqrt{p(\cdot)} \in \mathcal{Q}$ be a square-root density function with expansion satisfying*

$$q(\cdot) = \sum_{i=1}^\infty q_i \phi_i(\cdot) \,, \quad \text{and} \quad 1 = \int_\mathcal{D} q(x)^2 \, \mu(dx) = \sum_{i=1}^\infty q_i^2 \,, \tag{55}$$

*with random, real-valued coefficients $q_i, i = 1, \ldots, \infty$. Then, in the infinite-dimensional limit, spherical HMC follows the nonparametric Fisher metric's geodesic flows in the same way that Riemannian HMC follows the Fisher metric's geodesic flows over the parametric family of distributions $\mathcal{P}_\theta$.*

*Proof.* Each of these algorithms relies on a split Hamiltonian [36] integration scheme (e.g. Equation (36)), wherein the Hamiltonian of interest ($H$) is split into two Hamiltonians ($H^1 + H^2$) that are then iteratively simulated. The formal Hamiltonian for spherical HMC on $lim_{I \to \infty} \mathcal{S}^{I-1} = \mathcal{S}^\infty$ has the same form as in Equation (36), but in this case the velocity $v$ is restricted to the tangent space to $\mathcal{S}^\infty$ at $q$, $T_q\mathcal{S}^\infty$. The Hamiltonian corresponding to Riemannian HMC is also split in the following way [31]:

$$H(\theta, \xi) = -\log p(\theta) + \frac{1}{2}\log \mathcal{I}(\theta) + \frac{1}{2}\xi^T \mathcal{I}^{-1}(\theta)\xi \tag{56}$$

$$H^1(\theta, \xi) = -\log p(\theta) + \frac{1}{2}\log \mathcal{I}(\theta)$$

$$H^2(\theta, \xi) = \frac{1}{2}\xi^T \mathcal{I}^{-1}(\theta)\xi,$$

where $\mathcal{I}(\theta)$ is the Fisher information, and $\xi$ is the auxiliary momentum variable.

Switching out $\xi(t)$ for $\nabla_\theta \ell(\theta(t))$ in Equation (53), it follows that the solutions to the Hamilton's equations for Hamiltonian $H^2(\theta, \xi)$ are the geodesics on the Riemannian manifold $(\mathcal{P}_\theta, g_F)$. This is because the Hamiltonian flow $\theta(t)$ preserves $H^2(\theta, \xi)$:

$$\frac{d}{ds}E(\theta) = \frac{d}{ds}\frac{1}{2}\int_a^b \xi(t)^T \mathcal{I}(\theta(t))^{-1}\xi(t) = \frac{d}{ds}\frac{b-a}{2}\xi(a)^T \mathcal{I}(\theta(a))^{-1}\xi(a) = 0\,. \tag{57}$$

Thus, Riemannian HMC steps around the state space by minimizing the parametric Fisher energy.

In the same way, exchanging $v(t)$ for $\dot{q}(t)$ of Equation (54), it follows that the solutions to the Hamilton's equations for Hamiltonian $H^2(q, v)$ are geodesics on the Riemannian manifold $(\mathcal{S}^\infty, \langle \cdot, \cdot \rangle_{\ell^2})$ and, by Lemma 5, correspond to geodesics on $(\mathcal{P}, g_F(\cdot, \cdot))$. Hence, both formal algorithms move around the state space by iteratively perturbing the velocity ($H^1$) and travelling the geodesics corresponding to the parametric and nonparametric Fisher geometries, respectively.

Finally, Lemma 6 guarantees that the finite-dimensional spherical geodesics (used in practice) pass in the limit to the geodesics of the sphere $S^\infty$ and hence of $(\mathcal{P}, g_F(\cdot, \cdot))$. □

## 5 Empirical results

Here we apply the $\chi^2$-process density model to both simulated and real-world data. As stated in Section 3.1, the eigen-pairs corresponding to the GP with covariance operator (30) depend on both the dimension and the area of $\mathcal{D}$. When $\mathcal{D}$ is the one-dimensional unit interval, the eigen-pairs are given by

$$\lambda_i^2 = \sigma^2(\alpha + \pi^2 i^2)^{-s}, \quad \text{and} \quad \phi_i(x) = \sqrt{2}\cos(\pi i x)\,, \tag{58}$$

for $i \geq 0$. For $\mathcal{D}$ the two-dimensional unit square $\mathcal{D} = [0,1] \times [0,1]$, the eigen-pairs are given by

$$\lambda_i^2 = \sigma^2\left(\alpha + \pi^2(i_1^2 + i_2^2)\right)^{-s}, \quad \text{and} \quad \phi_i(x) = 2\cos(\pi i_1 x_1)\cos(\pi i_2 x_2)\,, \tag{59}$$

for $i_1, i_2 \geq 0$. See Beskos, Girolami, Lan, et al. [21] for a similar approach. In the following experiments, all Markov chains are initialized using Newton's method on the sphere (see Appendix A).

### 5.1 Simulated experiments

Figure 1 depicts 1,000 data points (red hash marks) drawn from four different beta distributions (density red) along with 100 MCMC draws from the posterior distribution based on the $\chi^2$-process density model. From left to right and top to bottom, the beta distribution parameters are $(1,1)$, $(5,2)$, $(.5,.5)$, and $(2,2)$. Note that while the individual posterior draws adhere closely to the sampled data, the variability in the posterior draws accounts for uncertainty and gives good coverage to the true density. The hyperparameter settings for the top-left plot is given by $(\sigma, \alpha, s) = (.5, 1, 1)$, and $(\sigma, \alpha, s) = (.5, .5, .8)$ is the hyperparameter setting for the rest. $I = 30$ for each example. 10,000 thinned MCMC iterations were used to make each figure.

Figure 2 depicts 1,000 data points (red) drawn from four different distributions on the unit square along with the contours of the pointwise median of 1,000 posterior draws from the $\chi^2$-process density model. The data in the first three plots was generated using truncated Gaussians and mixtures of truncated Gaussians. The data for the last plot was generated by Gaussian noise added to the uniform distribution on the circle. The model adapts easily to multimodal and patterned data samples. For all examples, the hyperparameters were fixed to $(\sigma, \alpha, s) = (.9, .1, 1.1)$. $0 \leq i_1, i_2 \leq 5$ for each example.

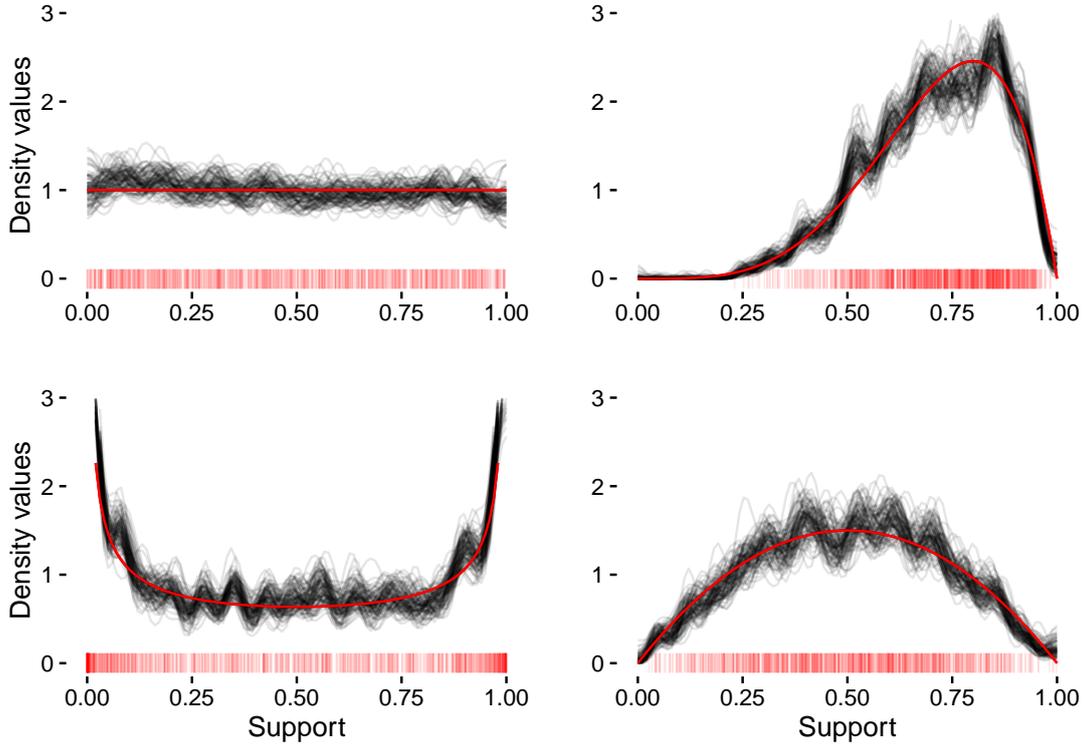

Figure 1: Each plot shows 100 posterior draws from the $\chi^2$-process density sampler. 1,000 data samples were drawn from a different beta distribution for each plot. The generating pdf is given in red, and the red hash marks describe the actual data produced.

## 5.2 Experiments with real-world data

Figure 3 features the British coal mine disaster data set, in which the dates of 191 disasters are recorded between the years of 1851 and 1967. In both plots, the dates are given in red. Two comparisons are implied by the figure. The first is a comparison between the variability of 100 posterior draws based on 191 data points (left plot) with the variability in 100 posterior draws based on 1,000 data points, as in Figure 1. One sees much less variability in the latter. The other comparison is between the close fit exhibited in the posterior draws of the left plot compared to the smooth fit shown by the pointwise quantiles (median, black; .25, blue; .75, blue). As we can see, our method is valid for modeling densities without periodic tendencies, despite the specific form of the basis. Both plots are based on 10,000 thinned MCMC iterations, with hyperparameter settings $(\sigma, \alpha, s) = (.5, .5, .8)$ with $I = 30$.

Figure 4 features Hutchings' bramble canes data (red) [38], [39], consisting of the locations of 823 bramble canes in a square plot. The left figure contains a heatmap of the pointwise posterior mean of the $\chi^2$-process density model, where black pertains to low density and white pertains to high density. Finally, a single contour (blue) at density level 0.3 divides the majority of points from areas of extremely low density. The hyperparameters were set to $(\sigma, \alpha, s) = (2, .01, 1.1)$ with $0 \leq i_1, i_2 \leq 5$, and the posterior sample featured 10,000 MCMC iterations. The right figure features 823 draws from the posterior predictive distribution of the $\chi^2$ process density model. Each draw from the posterior predictive distribution was obtained by randomly selecting one posterior draw from the $\chi^2$ process density model. Since this single posterior sample is itself a density function, one can then sample from its corresponding distribution using a rejection sampling scheme. There is a remarkable similarity between the posterior predictive sample (right, black) and the bramble canes data (left, red): despite a few differences, both low and high density regions are faithfully recovered.

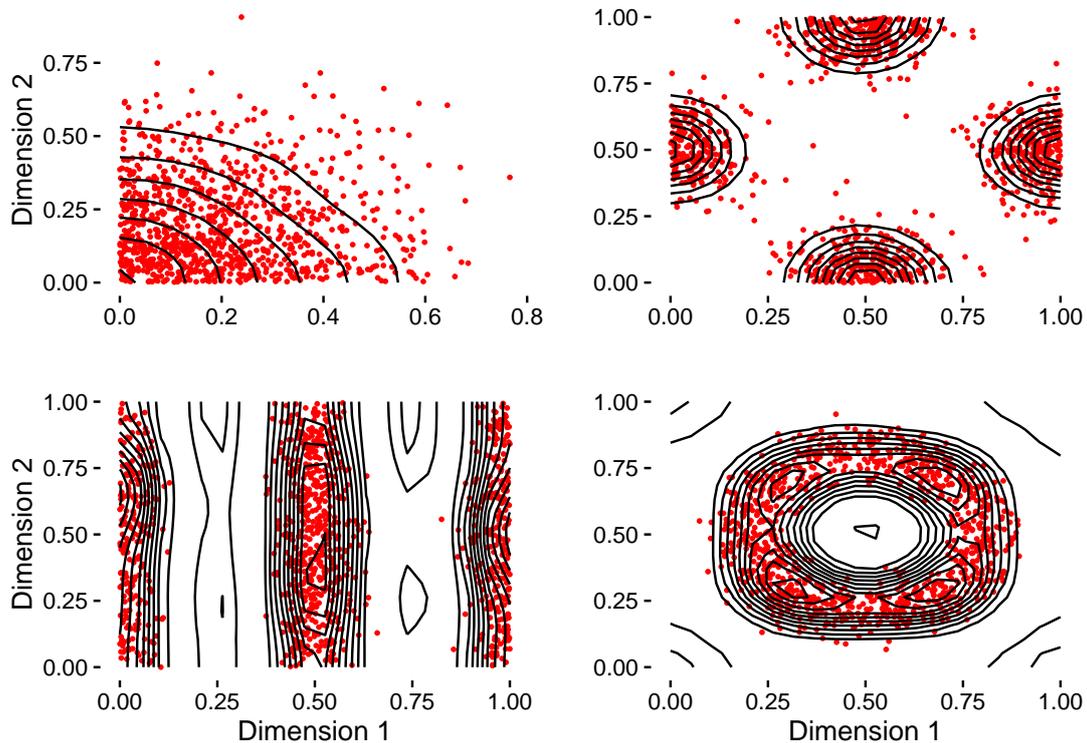

Figure 2: The contours (black) of the posterior median from 1,000 draws of the $\chi^2$-process density sampler. Each posterior is conditioned on 1,000 data points (red).

## 6 Discussion

The Fisher geometry is central to many areas of classical and parametric statistics. On the other hand, nonparametric methods—both Frequentist and Bayesian—is a vital area of statistical research with many realizations and applications. We presented a nonparametric extension to the parametric Fisher geometry and showed that this generalization is consistent with its parametric predecessor. To do so, the set of probability density functions over a given domain was defined to be an infinite-dimensional smooth manifold where each point is itself a density function. This manifold becomes a Riemannian manifold when equipped with the nonparametric Fisher information metric and is then identified with the infinite-dimensional sphere, a well understood geometric object for which results are readily obtainable. Indeed, the benefits of shifting focus to the infinite-dimensional sphere do not stop at theory. Due to the relationship between the nonparametric Fisher geometry and the infinite sphere, it proves convenient to define nonparametric models directly on this sphere.

We demonstrated one application of this approach in the form of Bayesian nonparametric density es-timation. The resulting $\chi^2$-process density model is flexible and computationally efficient: it is amenable to HMC and, in comparison to the cubic scaling of GP competitors, scales linearly in the number of data points. Of course, there is nothing *a priori* restricting the prior to be Gaussian [40], and an important next step is placing a prior on the number of basis functions to use, as is done in [20].

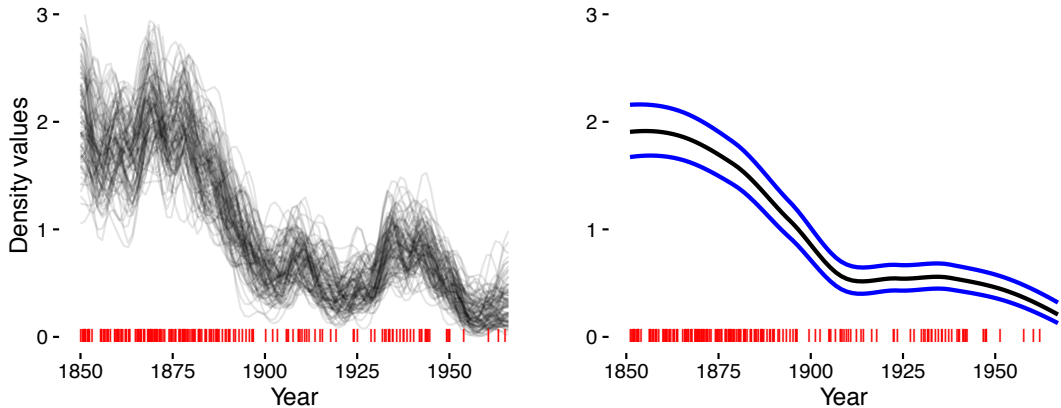

Figure 3: Coal mining disasters data: the left figure shows 100 posterior draws from the $\chi^2$-process density model (gray) over 191 vertical lines (red) marking the precise date of each disaster. The right figure shows the pointwise median (black) for the same sample as well as pointwise quantile bands (blue). Note how the undulations exhibited by individual draws does not appear in the quantile bands.

Moreover, spherical HMC uses geodesics to propose new states on the sphere, and these geodesic flows are formally equivalent to those derived on the $L^2$ sphere. Thus, the empirical effectiveness of spherical HMC in this context suggests that the proposals somehow adapt to the nonparametric Fisher geometry, and we showed that these proposals minimize the nonparametric Fisher energy in the same way that Riemanian HMC minimizes the parametric Fisher energy. We hope the $\chi^2$-process density model will serve to motivate extensions of HMC to Hilbert manifolds, of which the $L^2$ sphere is one example. It is now known how to perform HMC on certain finite dimensional manifolds [22], [31], [37] as well as Hilbert spaces [41]. We hope that the model presented here will motivate the extension of HMC technology to a large class of Hilbert manifolds, including the infinite-dimensional sphere.

The theoretical and methodological results presented in this paper are merely first steps in exploiting the simple geometry implied by the nonparametric Fisher metric. Whereas density estimation is perhaps the most obvious application, it is also one of the fundamental problems in statistics and thus has connections to many other areas of statistics and machine learning. On the other hand, methodologies such as functional regression and classification [42] can benefit from the use of random functions defined on the sphere, which objects we constructed and performed inference on. Additionally, the nonparametric methodology proposed in this paper was Bayesian, but the spherical representation of the nonparametric Fisher geometry has clear connections to Frequentist nonparametrics by way of the geometry of the bootstrap [43].

# Acknowledgement

AH is supported by NIH grant T32 AG000096. SL is supported by the DARPA funded program Enabling Quantification of Uncertainty in Physical Systems (EQUiPS), contract W911NF-15-2-0121. This work was partially supported by NIH grant R01-AI107034 and NSF grant DMS-1622490.

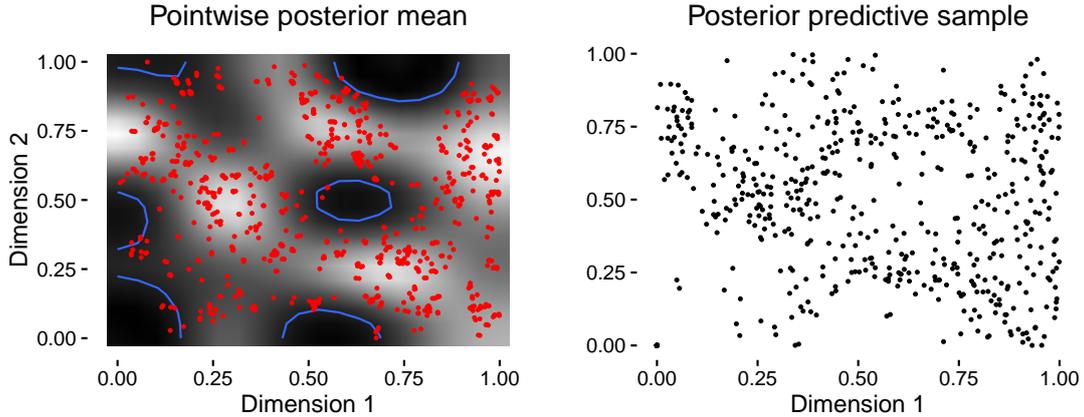

Figure 4: Hutchings' bramble canes data: the first figure depicts the 823 bramble canes (red), a heatmap of the pointwise posterior mean (black is low, white is high), and a single contour at density 0.3 (blue) including all but a few points. The second figure shows 823 draws from the $\chi^2$-process density posterior predictive distribution, obtained using a rejection sampling scheme.

## A  Initializing the Markov chain: Newton's method on the sphere

Starting with a Riemannian manifold $\mathcal{Q}$ isometrically embedded in Euclidean space, we consider function $F : \mathcal{Q} \to \mathbb{R}$.

**Definition 2.** *Given point $q_0 \in \mathcal{Q}$ and initial velocity $\dot{q}_0 \in T_{q_0}\mathcal{Q}$, we follow Edelman, Arias, and Smith [44] and define the Hessian of function $F$ along $\dot{q}_0$ as the matrix satisfying*

$$Hess\, F(\dot{q}_0, \dot{q}_0) = \frac{d^2}{dt^2}\Big|_{t=0} F(q(t)) . \tag{60}$$

**Proposition 1.** *Hess $F$ on the sphere is given by*

$$Hess\, F = F_{qq} - F_q^T q_0 \mathcal{I} , \tag{61}$$

*where $F_q$ and $F_{qq}$ are the Jacobian and usual Hessian matrices.*

*Proof.* We need the formula for the geodesic on the sphere given $q_0$ and $\dot{q}_0$. Letting $\alpha$ be the Euclidean norm of $\dot{q}_0$, the geodesic is given by:

$$q(t) = q_0 \cos(\alpha t) + \frac{\dot{q}_0}{\alpha} \sin(\alpha t) . \tag{62}$$

It is easy to verify that

$$\ddot{q}(t) = -\alpha^2\, q(t) . \tag{63}$$

Next the derivatives are given by:

$$\frac{d}{dt} F(q(t)) = \frac{\partial F}{\partial q}(y(t))\dot{q}(t) , \tag{64}$$

and

$$\frac{d^2}{dt^2} F(q(t)) = \dot{q}(t)^T \frac{\partial^2 F}{\partial q^2} \dot{q}(t) + \frac{\partial F}{\partial q}^T \ddot{q}(t) . \tag{65}$$

Combining (63) with (65) gives:

$$\frac{d^2}{dt^2} F(q(t)) = \dot{q}(t)^T \frac{\partial^2 F}{\partial q^2} \dot{q}(t) - \alpha^2 \frac{\partial F}{\partial q}^T q(t) \qquad (66)$$

$$= \dot{q}(t)^T \left( F_{qq} - F_q^T q(t) \mathcal{I} \right) \dot{q}(t).$$

Evaluating at $t = 0$ gives the result. □

Hess $F$ is the Hessian matrix at point $q_0$ in direction $\dot{q}_0$. Newton's method on the sphere is achieved by Algorithm 1.

---

**Algorithm 1** A single iteration of Newton's method on the sphere
---
1: Given point q on sphere:
2: Calculate $F_q$
3: Calculate Hess $F = F_{qq} - F_q^T q_0 \mathcal{I}$
4: Calculate $W = (\mathcal{I} - qq^T) \, Hess^{-1} F \, (\mathcal{I} - qq^T)$
5: $V \leftarrow -W F_q$
6: Progress along geodesic (62) with initial velocity $V$ for time 1.
7: $q \leftarrow q(1)$

---

## B  Relationship to the Cox process

The $\chi^2$-process density prior may be used to model the intensity function of a Cox process [45]. The Cox process is a point process over a given domain such that each realization at point $t$ is drawn from a Poisson distribution with intensity $\mu(s)$, where intensity function $\mu(\cdot)$ is itself a random process over the same given domain. Cox processes are useful for the analysis of spatial and time series data. Given $\mu(\cdot)$, the likelihood of such data $\{s_n\}_{n=1}^N$ is given by

$$p(\{s_n\}_{n=1}^N | \mu(\cdot)) = \exp\left(-\int_{\mathcal{D}} \mu(s)\, ds\right) \times \prod_{n=1}^N \mu(s_n). \qquad (67)$$

Bayesian inference on $\mu(\cdot)$ requires the calculation of two integrals, that over the parameter space and that from Equation (67). We make the latter integral trivial by modeling the intensity function as the product of a density function and a positively constrained random variable:

$$\mu(s) = M \times p(s) = M \times q(s)^2. \qquad (68)$$

In this case, the likelihood may be written

$$p(\{s_n\}_{n=1}^N | \mu(\cdot)) = \exp\left(-\int_{\mathcal{D}} M q(s)^2\, ds\right) \times \prod_{n=1}^N M q(s_n)^2 \qquad (69)$$

$$= \exp(-M)\, M^N \prod_{n=1}^N q(s_n)^2.$$

Since the likelihood factors in $M$ and $q(\cdot)$, it follows that the two random variables will be independent in posterior distribution *if* they are specified to be independent in prior distribution. Indeed, $M$ may even be given a conjugate prior: it is easy to see that

$$M \sim \Gamma(a, b), \quad \text{implies} \quad M | N \sim \Gamma(a + N, b + 1). \qquad (70)$$

Sampling from the joint posterior of $\mu(\cdot)$ is as simple as independently sampling $M$ from its posterior and $q^2(\cdot)$ from the $\chi^2$-process density sampler and then multiplying the two together. Such a model should be used with care. As a function of the data, the posterior distribution of $M$ solely depends on $N$, which is itself a single realization from a Poisson distribution. Thus, our $\chi^2$-process density prior–Cox process formulation is useful in situations where ample prior information on $M$ is available.